\documentclass[preprint]{acm_proc_article-sp}
\newcommand{\UPLB}{University of the Philippines Los Ba\~{n}os}

\begin{document}

\title{Preferential Attachment in\\an Internet-mediated Human Network}
\numberofauthors{1}
\author{
\alignauthor Chezka Camille P. Arevalo and Jaderick P. Pabico\\
   \affaddr{Institute of Computer Science}\\
   \affaddr{College of Arts and Sciences}\\
   \affaddr{\UPLB}\\
   \affaddr{College 4031, Laguna, Philippines}\\
   \affaddr{63-49-536-2313}\\
   \email{jppabico@uplb.edu.ph}
}
\date{}
\toappearbox{Contributed scientific paper to the 2009 Philippine Computing Science Congress, Silliman University, Dumaguete City, 2--3 March 2009.\\
This article reports an extension of previous presentations/publications in~\cite{pabico08a,pabico08b,pabico08c}.}

\maketitle

\begin{abstract}
In the advent of the Internet, web-mediated social networking has become of great influence to Filipinos. Networking sites such as Friendster, YouTube, FaceBook and MySpace are among the most well known sites on the Internet. These sites provide a wide range of services to users from different parts of the world, such as connecting and finding people, as well as, sharing and organizing contents. The popularity and accessibility of these sites enable information to be available. These allow people to analyze and study the characteristics of the population of online social networks. In this study, we developed a computer program to analyze the structural dynamics of a locally popular social networking site: The Friendster Network. Understanding the structural dynamics of a virtual community has many implications, such as finding an improvement on the current networking system, among others. Based on our analysis, we found out that users of the site exhibit preferential attachment to users with high number of friends. 
\end{abstract}

\section{Introduction}\label{sec:1}

%why is the problem interesting?
Preferential attachment is a process in which a quantity is distributed among a number of entities according to how much the entities already have, so that those entities which already have a lot of quantities receive more than those which have less~\cite{wikipedia}. In Internet-mediated human networks, such as those sites and services that are classified as social networks, the quantity distributed in preferential attachment is the number of relationship an entity has, while the entities are the site users themselves. Understanding the preferential attachment dynamics of Internet-mediated human networks has many uses. In the point of view of computing and information technology, understanding the structural dynamics of online social networks can help in improving the current systems. Similarly, it can also help in designing new applications for these systems and in understanding the impact of online social networks on the Internet. For instance, observing shared interest and trust of users can lead to algorithms which could give better results of the user's future searches. If future distributed online social networks become more popular and bandwidth-intensive, they can have a significant impact on Internet traffic, just as current peer-to-peer content distribution networks do~\cite{Ray}, allowing one to design a better network overlay system. 

Understanding the structural dynamics of social networks can also have an impact on the social science discipline. For instance, the information that can be gathered from the analysis can be used to test theories derived from the previous social studies conducted using small samples~\cite{Ahn}. Similarly, the results of such a study could also be used in the fields of information dissemination and mass communication. For example, politicians can use the knowledge for online campaign while the marketing industry people can use it for promoting products and companies. The reason for this is that new algorithms for determining authoritative sources in the web can be applied on social networks to determine influential users. Moreover, more ways on how to improve Internet
search, to filter email spam and understand how virus spreads, maybe contributed by such understanding. The knowledge will also play an important role in future online interaction and in locating and organizing information and knowledge. Thus, analyzing the structural dynamics of these social networks are of tremendous importance to social networking~\cite{Kanter}.

%what is the background of the previous solutions?
In our previous efforts, we used data mining and information theory techniques to extract and analyze on a community-scale the demography, friendship preferences, and network characteristics of a population, using as test bed the Friendster accounts of users whose listed hometown is Los Ba\~nos, Laguna~\cite{pabico08b,pabico08c}. The reason for this is that one of the most popular social networking sites among Filipinos is Friendster. An evidence of its popularity is the prevalent use of the street lingo "friendster" used by many Filipinos to refer to a friend. Friendster stores the participants' data such as gender, age, relationship status, geographic location, and list of friends, making it possible for an automated program to mine important data and relationships on a large scale. Based on this program, we found out about the Los Ba\~nos Friendster Network (LBFN) that:
\begin{enumerate}
\item There are more female users (52.34\%) than male (47.66\%);
\item Ages 15--25 of both genders compose 68\% of the users, with ages 26--40 following at 28\%, ages 41--85 at 4\%, and senior citizens (64--85 years old) at 1\%;
\item Homophily (i.e., birds-of-a-feather adage) is observed in the preferences of users with respect to age levels, such that they are strongly biased towards being friends with users of a similar age; 
\item There is heterophily in gender preference such that friendship among users of the opposite gender occurs more often.
\item The friendship network is well-connected and robust to node removal, such that users can still reach other users through another friend's circle of friends, even if another user leaves the network;
\item It exhibits a small-world characteristic with an average path length of 4.5 (maximum=12) among connected users, shorter than the well-known {\em six degrees of separation}~\cite{travers69}; And
\item The network exhibits a scale-free characteristics with heavily-tailed power-law distribution (with the power $\lambda = -1.02$ and $R^2 = 0.84$) suggesting the presence of many users acting as the network hubs.
\end{enumerate}

%what was done in this effort?
The data gathered from the previous study is based only on a static network created from one snapshot of the LBFN. For us to be able to understand the impact of users on the current underlying Internet overlay, we needed to analyze the network's dynamics over time. Thus, we extended our previous works~\cite{pabico08a,pabico08b,pabico08c} by capturing the structure of the LBFN over several snapshots.

%What will be presented in this paper?
In this paper, we will present the preferential attachment of LBFN users. We found out that users of the site exhibit preferential attachment to users with high number of friends. 

\section{Recent Selected Related Works on Online Social Networks}
During the time where the network of movie actors have been studied, people have already shown huge interest on the different structural properties such as degree distribution, scale-free and small-world characteristics of networks. This was followed by studies of different kinds of networks like the scientific collaboration and the human sexual contacts networks. However, the studies conducted were based on a small-scale analysis and it is said that the relationships between these kinds of networks differ from that of a normal friend relationship. Just recently, the number of online social networks has significantly increased making it possible to study huge social networks directly. However, it is observed that these huge networks' analysis are more focused on the cultural and business viewpoints only~\cite{Ahn}. 

There have already been previous studies related to online social networks. The first one was a study of four sites: Flickr, YouTube, Orkut and LiveJournal. The data set consisted of about 1.8 million users from Flickr, 5.2 million users from LiveJournal, 3 million users from Orkut, and 1.1 million users from YouTube. The study showed that the structure of social networks and its characteristics differ from those networks mentioned earlier. It was found that online social networks have more links and are highly clustered. Nodes with high number of links towards them also have a high probability of having a high number of links coming from them. These online social networks are composed of clusters which are highly connected. However, these clusters are composed of nodes with low number of links. This resulted to the inversely proportional values of the clustering coefficient with respect to the number of links of each node. Although the path lengths are short, most paths passed through nodes which are highly connected~\cite{Mislove}.

Another one investigated on the topological characteristics of huge online social networking services. The structures of three online social networking services, Cyworld, MySpace, and Orkut were compared. The number of examined users was 100,000 for each social networking site. Results showed that these networks follow the power-law distribution having a heavy tail. Based on the analysis of the degree distribution of Cyworld, researchers found out that it supported the claim that the diversity of the types of users greatly affects different network characteristics such as clustering coefficient, evolution of the network size, average path length and the network's diameter. The results of the analysis of MySpace and Orkut followed the patterns found in the different regions of the Cyworld network~\cite{Ahn}.

\section{Methodology}
\subsection{The Web Robot}
Instead of obtaining the data from the site operator, the website was crawled by accessing the public web interface provided. A spider-like computer program that "crawls" the Friendster website was developed that automatically visited the participants' web pages. 

To be able to view the profiles of other Friendster users, a person should be logged-in using a valid account. In relation to this, a new friendster account~$V$ referring to a real human Friendster user was created. It is assured that the data, which the web robot extracts, is from a real person since Friendster has filtered their database and prevented Pretenders, Fakesters and Fraudsters from intruding the network. The web robot was created using {\it Perl} scripts. Linux command line programs such as {\it grep} and {\it wget} were also utilized. The web robot uses the cookie file of the web browser where the user~$V$ is currently logged-in which makes it seem that the web robot is simply the user~$V$ visiting the profiles of other Friendster users~\cite{pabico08b}.

\subsection{Friendster Users}

The search tool provided by Friendster was used to extract the accounts of users whose listed hometown is Los Ba\~{n}os, Laguna. The search parameters that concerns the person's friendship preferences and relationship status were applied. The search tool produces an array of $p$~pages with $N$~unique accounts. The first $p-1$ pages contain 10 unique accounts each while the last page contains $N$~modulo 10 accounts. To be able to crawl the $p$~pages, a parameter is changed in each URL. The web robot extracted the account number, user name, age, gender, and relationship status of each user~\cite{pabico08b}. 

While crawling the web pages of each user, the list of friends of a participant were also extracted and crawled. The information gathered is stored in separate database tables named "account" and "friends". The first one having the demographic information of the participants which produces $N$~unique records corresponding to each Friendster account gained from the crawl. The other one takes note of the account number of the participants as well as the account number of his friends. The user's account number in the "account" table is used as a foreign key for the other table containing his friends~\cite{pabico08b}.

\subsection{Creating and Analyzing the LBFN}

The friendship network was created using the data in the table "friends". Each account was treated as a vertex while the relationship between accounts as edges. From these, a $N \times N$ adjacency matrix was created wherein the value of the element is~1 if a relationship between users~$i$ and~$j$ exists in the table, otherwise, the value is~0.
	
With the help of Pajek, a tool for analyzing and drawing graphs of large networks ~\cite{Pajek}, the following network metrics were computed:
\begin{enumerate}
\item Degree distribution - It is the probability distribution of the number of connections of a node with respect to other nodes. Networks which follows the power-law distribution having a heavy tail is considered as scale-free~\cite{Barabasi}. 
\item Average Separation of Members - This is the average number of friends along the shortest paths over all pairs in which a person can reach another person. It shows the network's interconnectedness~\cite{Barabasi}. 
\item Clustering Coefficient - It tells how well connected a participant's friends are. It is the probability that a person's friends are also friends~\cite{Barabasi}.
\item Size of the Largest Cluster - In this case, this is defined as the highest number of links derived from the node with the highest number of friends.
\item Average Degree - The average degree can also be referred to as the average number of friends of a participant. This is computed by summing up the total number of friends of a person divided by the total number of participants involved in the network.
\item Preferential Attachment - This is the behavior wherein there is a high probability that a new node is more likely to connect to nodes which already have a high number of links to other nodes~\cite{Barabasi}.
\end{enumerate}

Three snapshots of LBFN were taken on August 5, August 26, and September 2, 2008.

\section{Results and Discussion}

\begin{figure}[ht]
\centering
\epsfig{file=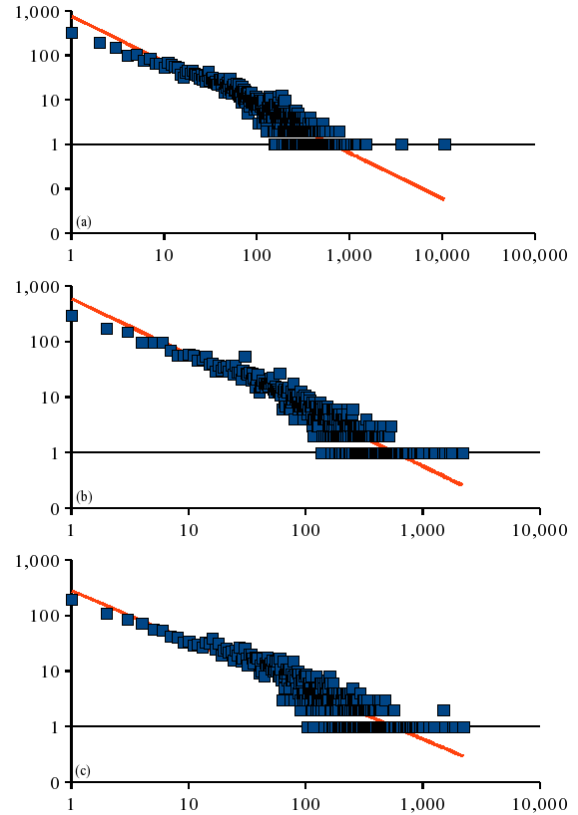, width=3in}
\caption{Log-log plot of the number of friends $\times$ frequency obeys the power law distribution over different snapshots: (a) August 5 (b) August 26 (c) September 2. Lines on each plot is the power-law fit using regression analysis.}
\label{fig1}
\end{figure}

Figure~\ref{fig1} shows the distribution of the number of friends in the log-log scale taken from each LBFN snapshot. The respective degree distributions indicate that the LBFN continues to be scale-free with a power-law tail over time. 
	 
\begin{figure}[ht]
\centering
\epsfig{file=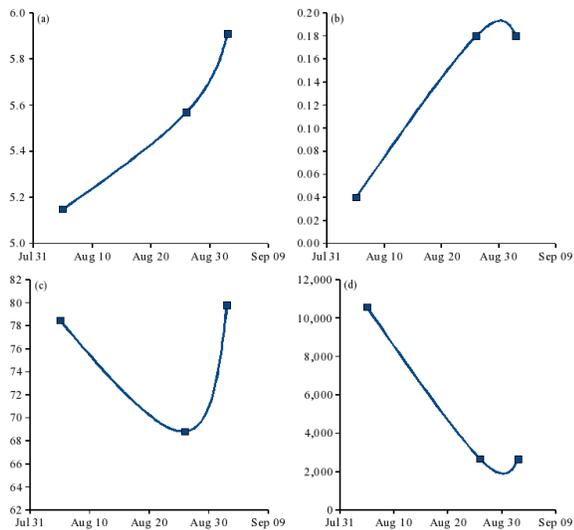, width=3in}
\caption{Graphs of (a) the average separation of nodes, (b) clustering coefficient, (c) frequency of the average number of friends of a person, and (d) largest cluster of the networks over time.}
\label{fig2}
\end{figure}

Figure~\ref{fig2}a shows that the average separation of nodes in LBFN increases in time. This means that through time, a person can be reached by another person through a friend of a friend at a much longer path or at a higher number of persons. We can only speculate a reason for this phenomenon: with the occurrence of new members, only few links are added to the network, which resulted to a larger network diameter.

The clustering coefficient for the networks as a function of time is shown in Figure~\ref{fig2}b. The results show its agreement with the separation measurements mentioned above. The values of the clustering coefficient range from 0.0352 to 0.1824 which suggest a weak interconnectedness. This means that there is a low probability that a person's friends are also connected to each other.

In Figure~\ref{fig2}d, the trend for the relative size of the largest cluster is shown. It is evident in the figure that the size of the largest cluster decreases. One possible reason is that the availability of accounts in Friendster is becoming less through time, with some of the accounts going private and unreadable to the web robot. This is based on the observation that the size of the networks decreases starting from the first network snapshot. It is possible that the largest cluster in the previous snapshot has already become private at the time. Friendster is also a dynamic network wherein a user can delete a friend, decreasing the size of the largest cluster.

\begin{figure}[ht]
\centering
\epsfig{file=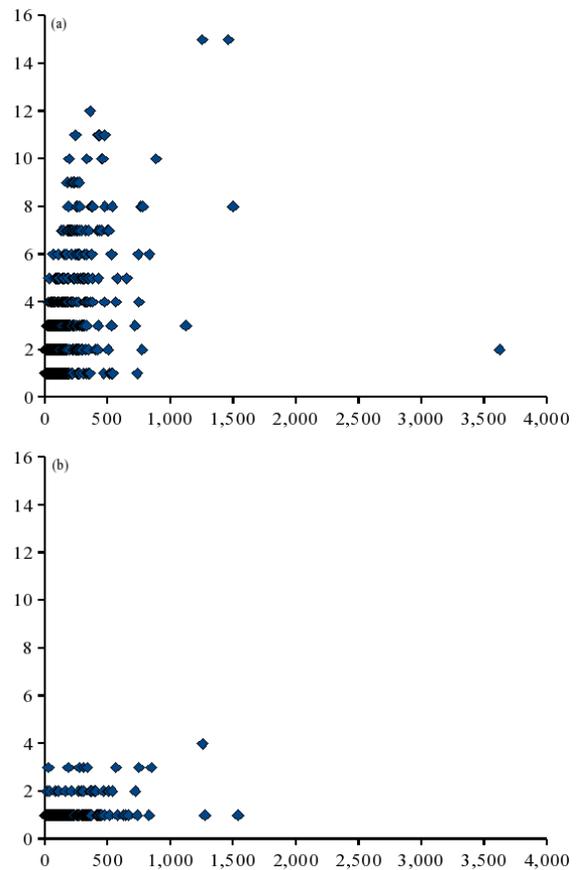, width=3in}
\caption{Graph of preferential attachment of nodes, node degree $\times$ new links, (a) from August 5 to August 26 (b) from August 26 to September 2.}\label{fig5}
\end{figure}

Figure~\ref{fig5} shows the trend of the number of new links to old nodes. Results show that more new nodes attach to nodes which have a high number of existing links. Users tend to be friends with those who already have a large number of friends suggesting that there is preferential attachment in the LBFN. This means that there is higher probability that a person~$A$ is connected to a person~$B$ where $B$~has a relatively large number of friends or links.

\section{Conclusion}

This study presents new results from an extension of previous studies. The dynamics of LBFN were measured using a web robot that we developed. Based on the analysis, the following results were found:

\begin{enumerate}
\item New users exhibit preferential attachment to users with high number of friends;
\item The average separation increases over time, suggesting that the interconnectedness of the users are getting weaker;
\item The largest cluster decreases through time; And
\item The average number of friends decreases through time which shows that, on the average, users lose more friends than acquire new ones.
\end{enumerate}

\section{Acknowledgments}

The authors thank the Institute of Computer Science and the College of Arts and Sciences, University of the Philippines Los Ba\~{n}os for its financial support of this work through CAS-TF \#8217300 and ICS-GF \#2326103, respectively.

\bibliographystyle{plain}
\bibliography{pref-attach}

\begin{thebibliography}{10}

\bibitem{Ahn}
Y.Y. Ahn, S.~Han, H.~Kwak, S.~Moon, and H.~Jeong.
\newblock Analysis of topological characteristics of huge online social
  networking services.
\newblock In {\em Proceedings of the 16th International on World Wide Web
  Conference (WWW'07)}, pages 8--12,, Banff, Canada, 2007.

\bibitem{pabico08c}
C.C.P. Arevalo and J.P. Pabico.
\newblock Automatic characterization of a friendster network using a data
  mining webbot.
\newblock In {\em Proceedings (CDROM) of the 4th Network of CALABARZON
  Educational Institutions, Inc. (NOCEI) Research Forum}, 2008.

\bibitem{Barabasi}
A.L. Barabasi, H.~Jeong, Z.~Neda, E.~Ravasz, A.~Schubert, and T.~Vicsek.
\newblock Evolution of the social network of scientific collaborations.
\newblock {\em Physica A}, 311:590--614, 2002.

\bibitem{Kanter}
B.~Kanter.
\newblock {Determining Your Social Network Needs}, 2008.
\newblock http://www.techsoup.org/.

\bibitem{Mislove}
A.~Mislove, M.~Marcon, K.~Gummadi, P.~Druschel, and B.~Bhattacharjee.
\newblock Measurement and analysis of online social networks.
\newblock In {\em Proceedings of Internet Measurement Conference (IMC'07)},
  pages 24--26,, San Diego, California, USA, 2007.

\bibitem{pabico08b}
J.P. Pabico.
\newblock Inferences in a virtual community: Demography, user preferences and
  network topology.
\newblock {\em Philippine Information Technology Journal}, 1(2):2--8, 2008.

\bibitem{pabico08a}
J.P. Pabico and C.C.P. Arevalo.
\newblock Patterns of internet-based friendship among residents of los
  ba\~{n}os laguna: The friendster case.
\newblock {\em Transactions of the National Academy of Science and Technology
  of the Philippines}, 30(1):220, 2008.

\bibitem{Pajek}
{Networks/Pajek: Program for Large Network Analysis}, 2008.
\newblock http://vlado.fmf.uni-lj.si/pub/networks/pajek.

\bibitem{Ray}
S.~Ray.
\newblock {The Importance of Social Networking}, 2007.
\newblock http://ezinearticles.com/.

\bibitem{travers69}
J.~Travers and S.~Milgram.
\newblock An experimental study of the small world problem.
\newblock {\em Sociometry}, 32(4):425--443, 1969.

\bibitem{wikipedia}
wikipedia.org.
\newblock {Preferential Attachment}, 2008.
\newblock http://en.wikipedia.org.

\end{thebibliography}
\end{document}